\documentclass[aps,prd,showpacs,showkeys,superscriptaddress]{revtex4}
\usepackage{graphics}
\def\ntext{\begin{multicols}{2}\global\columnwidth20.5pc}
\def\wtext{\end{multicols} \global\columnwidth42.5pc}

\def\be{\begin{equation}}
\def\ee{\end{equation}}
\def\bea{\begin{eqnarray}}
\def\eea{\end{eqnarray}}
\def\bey{\begin{eqnarray*}}
\def\eey{\end{eqnarray*}}
\def\bml{\begin{mathletters}}
\def\eml{\end{mathletters}}
\def\ba{\begin{array}}
\def\ea{\end{array}}

\def\nn{\nonumber}
\def\to{\rightarrow}
\def\<{\left\langle}
\def\>{\right\rangle}
\def\({\left(}
\def\){\right)}
\def\sumint{\sum\!\!\!\!\!\!\!\!\int}
\def\Li{\,{\rm Li}}
\def\del{\partial}
\def\bm{\bar{\mu}}
\def\e{{\rm e}}
\def\a{\alpha}

\def\C{\hat{C}}
\def\J{\hat{J}}
\def\M{\hat{M}}
\def\B{\hat{B}}

\def\U{{\rm U}}
\def\SU{{\rm SU}}
\def\usp{{\rm usp}}
\def\USp{{\rm USp}}
\def\tr{\,{\rm tr}\,}

\def\Psit{\Psi^{T}}
\def\psib{\bar{\psi}}
\def\psid{\psi^{\dagger}}
\def\psis{\psi^{*}}

\def\chib{\bar{\chi}}
\def\chid{\chi^{\dagger}}
\def\chis{\chi^{*}}

\def\D{D\!\!\!\!\slash\;}

\def\Q{\tilde{Q}}

\begin{document}
\preprint{}
\title{3D two-color QCD at finite temperature and baryon density}
\author{Gerald V. Dunne}
\affiliation{Department of Physics, University of Connecticut,
Storrs, CT 06269-3046, USA}
\author{Shinsuke M. Nishigaki}
\affiliation{Department of Physics, University of Connecticut,
Storrs, CT 06269-3046, USA}
\affiliation{
Department of Material Science,
Faculty of Science \& Engineering,
Shimane University,
Matsue 690-8504, Japan
}
\date{\today}
\begin{abstract}
We study the phase diagram for two-color QCD in three-dimensional spacetime,
as a function of temperature and baryon chemical potential, using the 
low-energy effective Lagrangian approach. We show one-loop 
renormalizability at zero temperature, and then use the one-loop 
effective Lagrangian at finite temperature and chemical potential to 
show that at low temperature there is a critical line separating the 
normal and diquark phase, with this critical line ending at a tricritical 
point. This phase structure is qualitatively similar to that found 
recently by Splittorff et al.\ for two-color QCD in four-dimensional 
spacetime, although the details are quite different, due to the 
different symmetries and the different loop and infrared properties of 
three-dimensional spacetime.
\end{abstract}
\maketitle

\section{introduction}
A systematic approach 
to the investigation of the symmetry breaking structure of fermion-gauge 
theories is to use low-energy effective Lagrangians 
\cite{gasser,weinberg} to study the Goldstone modes corresponding to the 
spontaneously broken global symmetries \cite{jac1}.  Ideally, one would 
like to understand the phase diagram of fermion-gauge theories in the 
$(T, \mu)$ plane, where $T$ is the temperature, and $\mu$ is the (baryon 
number) chemical potential. However, it is not known how to analyze 
general fermion-gauge systems at finite baryon density, since the baryon 
number chemical potential makes the Euclidean Dirac operator 
non-Hermitian and the Boltzmann weight complex. This problem can be 
avoided, as advocated in \cite{hands,kst,kogut,jac2,jac3}, by 
considering ``two-color QCD'', for which the fundamental representation 
of $\SU(2)$ is pseudoreal. {}From studies of 4D theories, it has long been 
appreciated that the $\SU(2)$ theory exhibits exotic types of 
spontaneous breakdown of global symmetry  \cite{peskin}. Quarks and 
charge-conjugated antiquarks are combined into an extended flavor 
multiplet, which is expected to break into its extended vector subgroup. 
This has the consequence that in lattice simulations the Boltzmann 
weight is real
even at finite (baryon number) chemical potential $\mu$.
For an even number of flavors, the Boltzmann weight is also positive definite.
Thus, analytic predictions can be quantitatively compared with Monte Carlo
simulations in lattice gauge theory \cite{dagotto,baillie}, provided the 
lattice regularization respects the relevant flavor symmetry group. Such 
studies have recently been carried out \cite{kogut2} in 4D QCD with 
quarks in pseudoreal (and real) representations at
finite $\mu$.

In \cite{dn} we studied the $T=0$ case of 3D two-color QCD, and found 
that the phase structure predicted by the tree-level effective potential 
was qualitatively the same as the 4D case studied in 
\cite{kst,kogut,stv1}, even though the details of the symmetry breaking 
in 3D are completely different from in 4D. This universality between 3D 
and 4D corresponds to the existence of two different breakings of the 
original global symmetry by a mass and a chemical potential term. In 
this paper we extend the 3D analysis of \cite{dn} to include nonzero 
temperature effects via the one-loop effective potential, as was done 
recently for the 4D system in \cite{stv2}.  The loop effects introduce 
another difference between 3D and 4D, as infrared physics is more 
significant in 3D. As a result, we find some similarities with the 4D 
case, but also some important differences.

In 3D, with an {\it even} number $N_F$ of flavors of massless complex 
fermions,
denoted by the $N_F/2$ pairs $\psi_f, \chi_f$,
one can predict spontaneous flavor symmetry breaking patterns along 
similar lines to 4D QCD. In 3D the generic situation is that the 
continuous part of the global symmetry group is broken
according to \cite{pisarski1,poly}
\be
\U(N_F) \rightarrow \U(N_F/2)\times \U(N_F/2)
\label{break2}
\ee
by the quark-antiquark condensate
\be
\sum_{f=1}^{N_F/2}
\left(
\langle\bar{\psi}_f  {\psi}_f \rangle-
\langle\bar{\chi}_f  {\chi}_f \rangle \right),
\label{mcondensate}
\ee
Evidence for such a symmetry breaking pattern has been observed in 3D 
lattice simulations  \cite{damgaard} with gauge group $\SU(3)$.  This 
pattern of flavor symmetry breaking can
also be predicted for 3D QCD at large $N_{C}$ using the Coleman--Witten 
argument \cite{CW}. The 3D symmetry breaking pattern in (\ref{break2}) is
for quarks in a complex representation of the gauge group,
and is expected to apply to a $\U(1)$ theory and to $\SU(N_C)$ theories 
with number of colors $N_C\geq 3$. For $\SU(2)$, with fundamental 
quarks, the symmetry breaking pattern is expected to be different again, 
due to the pseudoreality of the fundamental representation.  The 
pseudoreality of the fundamental representation of $\SU(2)$ means that 
the naive $\U(N_F)$ flavor symmetry is extended to $\USp(2N_F)$, and the 
continuous part of this global symmetry group is predicted
to break down in 3D as \cite{magnea}
\be
\USp(2N_F) \rightarrow \USp(N_F)\times \USp(N_F).
\label{break1}
\ee
This is different from the generic 3D symmetry breaking pattern in 
(\ref{break2}), and also is different from 4D theories where the standard flavor symmetry breaking 
patterns are
\begin{eqnarray}
\U(N_F)_L\times \U(N_F)_R \to \SU(N_F)_V &,& {\rm for}\,\, \SU(N_C\geq 
3)\nonumber\\
\U(2N_F)\to \USp(2N_F)   &,& {\rm for}\,\, \SU(2)\, .
\label{4d}
\end{eqnarray}
In 4D the $\U(N_F)$ flavor symmetries are first broken to $\SU(N_F)$ 
symmetries by the axial anomaly, and then broken by the chiral 
condensate, with the net breakings as shown in (\ref{4d}). Physically, 
the differences between the 3D and 4D cases reflect the differences 
between the anomalous discrete symmetries of parity and chirality in 3D 
and 4D, respectively.

In section II we review the global symmetries and their cosets that are 
relevant for the low energy effective Lagrangian description of the 3D 
two-color QCD system, with quarks in the fundamental representation.  In 
section III we compute the one-loop effective action at finite 
temperature and finite baryon chemical potential. This result is used in 
section IV for a Landau-Ginzburg description of the phase diagram, using 
the rotation angle $\alpha$ which characterizes the tree-level vacuum 
state as an order parameter. Section V contains our conclusions and some 
discussion.

\section{Symmetries and Effective Lagrangian}
\subsection{Enlarged flavor symmetry}
The fermionic part of the microscopic Lagrangian for two-color QCD
with $N_{F}=2n$ flavors of quarks in Euclidean 3D space is
\be
{\cal L}_f=\psib\D\psi+\chib\D \chi+m(\psib\psi-\chib\chi) - \mu 
(\psi^\dagger\psi+
\chi^\dagger\chi).
\label{microlag}
\ee
Here $\psi=\psi_{f}^{i}$, $\psid=\psis{}_{f}^{i}$,
$\chi=\chi_{f}^{i}$, $\chid=\chis{}_{f}^{i}$ are independent
two-component spinor fields,
with
the color index $i=1, 2$ and flavor index
$f=1,\ldots,n$ being suppressed.
Pauli matrices denoted as $\sigma_\nu$, with $\nu=1,2,3,$ are employed
to represent the Euclidean Dirac matrix algebra, and those
denoted as $\tau_{\alpha}$, with $\alpha=1,2,3$, are employed
to represent the gauge group algebra.
We choose $x_3$ to be the Euclidean time direction and define
$\psib=\psid \sigma_{3}$, $\chib=\chid \sigma_{3}$.
The Dirac operator is $\D=\sigma_\nu D_\nu$, and
the covariant derivative is
$D_\nu=\partial_\nu+i A_\nu$,
and the gauge field $A_\nu=A_\nu^\alpha \tau_\alpha$
is Hermitian and ${\rm su}(2)$ valued.

Due to the pseudoreality of the $\SU(2)$ Dirac operator,
the kinetic part of the Lagrangian (\ref{microlag}) is invariant under a 
symmetry group
larger than the apparent $\U(2n)$ symmetry. One can put $\psi$, $\chi$,
$\tilde{\psi}$, and $\tilde{\chi}$ into a single flavor $4n$-plet,
\be
\Psi=\left[\ba{cc}\psi\\
\chi\\
\sigma_{2}\tau_2\psib^T\\
\sigma_{2}\tau_2\chib^T
\ea\right] ,
\label{4nplet}
\ee
so that
\be
{\cal L}_f=\frac12 \Psit\sigma_{2}\tau_2\D \hat{I} \Psi
+\frac{m}{2} \Psit\sigma_{2}\tau_2 \hat{M} \Psi
-\frac{\mu}{2} \Psi^T i\sigma_1 \tau_2 \hat{C} \Psi
\label{microlag2}
\ee
where the $4n\times 4n$ matrices $\hat{I}$, $\hat{M}$ and $\hat{C}$ are
\be
\hat{I}= \left[\ba{c|c}& -\openone_{2n}\\ \hline\openone_{2n}& 
\ea\right],\ \ 
\hat{M}= \left[\ba{cc|cc}
& & -\openone_{n}&\\
& & & \openone_{n}\\
\hline
\openone_{n}&&&\\
& -\openone_{n}&&
\ea\right], \ \ 
\hat{C}= \left[\ba{c|c}
& \openone_{2n}\\
\hline
\openone_{2n}&
\ea\right] .
\label{matrices}
\ee
The kinetic term in the microscopic Lagrangian (\ref{microlag2}) has 
extended flavor symmetry group being the unitary symplectic group 
$\USp(4n)$,
\be
\Psi\to S\Psi, \ \ S^T \hat{I} S=\hat{I} ,\ \ S^\dagger S=\openone_{4n}.
\ee
The above extension of the flavor symmetry group is analogous to what 
happens in two-color QCD in 4D, where the conjugated right-handed spinor 
$\sigma_{2}\tau_2\psi_R^*$ transforms as the left-handed spinor $\psi_L$ 
does under gauge and Lorentz transformations, so that the chiral 
$\SU(N_F)_L\times \SU(N_F)_R$ symmetry is extended to $\SU(2N_F)$.

In the 3D case, the flavor group $\USp(4n)$ is broken down to 
$\USp(2n)\times\USp(2n)$
explicitly by the mass term in (\ref{microlag2}), or spontaneously by 
the quark-antiquark condensate (\ref{mcondensate}), if formed.
In the latter case, the Goldstone manifold is thus  given
by the coset space
$\USp(4n)/(\USp(2n)\times\USp(2n))$, which has $4n^2$ independent
degrees of freedom, and can be parametrized as
\be
\Sigma=S \Sigma_c S^T,\ \ \ \Sigma_c=\hat{M}^\dagger,
\label{Sigmadef}
\ee
where
\be
S(x)=\exp\left(\frac{i \Pi(x)}{2F}\right),\ \ \
\Pi(x)=\pi_a(x) X_a.
\label{S}
\ee
The fields $\pi_a$ are the Goldstone modes, and
the $4n^2$ generators $X_a$ span the subspace
$\usp(4n)-(\usp(2n)\oplus\usp(2n))$.
The choice of $\Sigma_c=\M^\dagger$  leads to the block representation
of the generators $X_a$ as
\be
\Pi=
\frac12
\left[
\ba{cc|cc}
& P &  & Q \\
P^\dagger &  & Q^T &  \\
\hline
& Q^* &  & -P^* \\
Q^\dagger &  & -P^T &
\ea
\right] ,
\label{Pi}
\ee
where $P$ and $Q$ are $n\times n$ complex matrices,
each having $2n^2$ degrees of freedom.
The chemical potential term in (\ref{microlag2})
explicitly breaks the extended flavor group $\USp(4n)$ down to its
unextended $\U(2n)$ subgroup, while in the presence of {\em both}
mass and chemical potential terms,
the surviving global symmetry becomes $\U(n)\times\U(n)$,  which is the 
intersection of
$\USp(2n)\times \USp(2n)$ and $\U(2n)$.

\subsection{Low-energy effective Lagrangian}

Coupling the fermionic  Lagrangian (\ref{microlag2}) to the $\SU(2)$ 
gauge field
gives the the microscopic Lagrangian
\bea
{\cal L}&=&
-\frac{1}{2g^2}\tr F_{\mu\nu}F_{\mu\nu}+
{\cal L}_f .
\label{LQCD}
\eea
In the low energy ($\ll\Lambda_{\rm QCD_3}$) regime,
where fundamental particles are confined and
Goldstone bosons dominate, we can define a low energy effective theory.
The kinetic term of the effective Lagrangian
describing the Goldstone modes $\Sigma$, parametrized as in 
(\ref{Sigmadef}) and (\ref{S}),
should be invariant under the action of the global $\USp(4n)$ group
\be
\Sigma(x)\to s \Sigma(x) s^T, \quad s\in \USp(4n)
\ee
with $s$ in the antisymmetric tensor representation. Invoking the 
standard symmetry principles of chiral perturbation theory, and the 
principle of local flavor symmetry, the effective Lagrangian  is
\bea
L=\frac{F^2}{2}
\tr \nabla_\nu \Sigma\nabla_\nu \Sigma^\dagger -F^2 M^2 \tr (\M \Sigma)
\label{efflag}
\eea
where the flavor-covariant derivatives are \cite{kst}
\bea
\nabla_\nu \Sigma&=&
\partial_\nu \Sigma-
\mu(\Sigma B_\nu^\dagger + B_\nu^\dagger \Sigma)\nn\\
\nabla_\nu \Sigma^\dagger&=&
\partial_\nu \Sigma^\dagger+
\mu(\Sigma^\dagger  B_\nu + B_\nu  \Sigma^\dagger).
\eea
with
\bea
B_\nu=\B \delta_{\nu,3}  ,\ \
\B=\C I=
\left[
\ba{c|c}
\openone_{2n} & \\
\hline
& - \openone_{2n}
\ea
\right] .
\label{B}
\eea

\subsection{Phases at tree level}
At zero temperature, the static part of the effective Lagrangian 
(\ref{efflag}),
i.e.\ the effective potential, determines the vacuum alignment of the 
field $\Sigma$.
Introducing a dimensionless parameter $\xi=2\mu/M$ to represent the chemical
potential, we find
\be
L_{\rm st}(\Sigma)={F^2 M^2}{}\left(
-\tr (\M \Sigma)
-\frac{\xi^2}{4} \tr( \B\Sigma\B\Sigma^\dagger )- n\xi^2\right).
\label{Veff}
\ee
The above two terms in $\Sigma$ compete for the direction of the condensate
which we denote by $\bar{\Sigma}$. The condensate that gives the global 
minimum of the static tree level effective Lagrangian is
\be
\bar{\Sigma}_\alpha=\M^\dagger \cos \alpha + \J \sin \alpha,
\label{bSigma}
\ee
which is parametrized by $\alpha$ where \be
\cos\alpha=\min ( 1,  \xi^{-2} ) .
\label{cosr}
\ee

\subsubsection{Normal phase}

When $\xi<1$, ({\it i.e.}, $\mu<\frac{M}{2}$), the vacuum orientation of 
the condensate
does not depend on $\xi$ and is given by $\bar{\Sigma}=\M^\dagger$.
Expanding $\Sigma$ around $\M^\dagger$
using the Goldstone field defined in (\ref{S}) and (\ref{Pi}) we find
\be
L=L_{\rm st}(\bar{\Sigma}_\alpha)+
\frac12\tr  \partial_\nu P^\dagger  \partial_\nu P
+\frac{M^2}{2} \tr  P^\dagger   P
+\frac12 \tr \partial_\nu Q^\dagger  \partial_\nu Q
- 2\mu \tr Q^\dagger  \partial_3 Q
+\Bigl( \frac{M^2}{2}-2\mu^2 \Bigr) \tr  Q^\dagger   Q
+\cdots .
\label{Lquad}
\ee
{}From this we can identify the excitations. The baryon charge $b$ of the 
excitations are :
$b=0$  for $P$ and $P^\dagger$;
$b=2$ for $Q$;  and
$b=-2$ for $Q^\dagger$.
The pole masses for these excitations are
\bea
M &{\rm for}& P\,\, {\rm and}\,\, P^\dagger\nn\\
M-2\mu &{\rm for}& Q\nn\\
M+2\mu &{\rm for}& Q^\dagger \, .
\eea
Each mode belongs to a representation of dimension $n^2$.

\subsubsection{Diquark condensation phase}

In the regime $\xi>1$, ({\it i.e.}, $\mu > \frac{M}{2}$),  the vacuum 
condensate has $\U(n)$ degeneracy corresponding to
$n^2$ true Goldstone modes. This change of massless modes indicates a 
second-order phase transition at $\xi=1$. When $\xi>1$ the configuration 
(\ref{bSigma})
begins to rotate away from $\hat{M}^\dagger$ according to $\cos \alpha = 
\xi^{-2}$.
This rotation can  be written as
\be
\bar{\Sigma}_\alpha= s_\alpha \M^\dagger s_\alpha^T = s_\alpha^2 
\M^\dagger,\ \
s_\alpha = \e^{i \frac{\alpha}{2} X_2},
\ee
where
$X_2$ is the generator  that rotates $\M$ into $\J$ .
We parametrize the fluctuation
around the vacuum $\bar{\Sigma}_\alpha$ as
\be
\Sigma=s_\alpha S \M^\dagger  S^T  s_\alpha^T,
\label{sSMSs}
\ee
where $S$ are generated by the unrotated generators $X_a$.

In this phase we expand around the rotated value of the condensate
$\bar{\Sigma}_\alpha$, as in (\ref{sSMSs}).
To second order
\bea
L&=&L_{\rm st}(\bar{\Sigma}_0)
+\frac12\tr  \partial_\nu P_S^\dagger  \partial_\nu P_S
+\frac12\tr  \partial_\nu P_A^\dagger  \partial_\nu P_A
+\frac12 \tr (\partial_\nu Q_R)^2
+\frac12 \tr (\partial_\nu Q_I)^2
\nn\\
&+&
i M \xi^{-1}
\tr
(Q_R^\dagger \partial_3 Q_I
-Q_I^\dagger \partial_3 Q_R)
+
\frac{M^2}{2}\left[
\xi^2 \tr P_S P_S^\dagger
+\xi^{-2} \tr P_A P_A^\dagger
+(\xi^2-\xi^{-2}) \tr Q_I^2  \right]
+\cdots .
\label{Lquaddq}
\eea
Here $P_{S,A}=(P\pm P^T)/2$, and $Q_{R,I}$ are the real and imaginary 
parts of the complex matrix $Q$. The  $P_{S,A}$ fields have conventional 
dispersion relations with masses
\bea
2\mu &{\rm for}& P_S\nn\\
\frac{M^2}{2\mu}&{\rm for}&  P_A
\eea
and lie in representations of dimension $n(n+1)$ and $n(n-1)$, respectively.

On the other hand, the dispersion laws for the $Q$ fields are 
unconventional due to the linear time derivative terms in 
(\ref{Lquaddq}), and are determined by
the secular equation
\be
\det
\left[
\ba{cc}
[E^2-{\bf p}^2]  &
2iE M \xi^{-1} \\
-2iE M \xi^{-1}&
[E^2-{\bf p}^2  -M^2(\xi^2-\xi^{-2})]
\ea
\right]
=0.
\label{secular}
\ee
Diagonalizing, we find one true Goldstone field, which we denote 
$\tilde{Q}$, and its orthogonal complement denoted $\tilde{Q}^\dagger$. 
These excitations have masses
\bea
0&{\rm for}& \Q\nn\\
2\mu\sqrt{1+3\left( \frac{M}{2\mu}\right)^4}&{\rm for}&\Q^\dagger
\eea
and each lies in a representation of dimension $n^2$.
Note that fields with this unconventional form of dispersion also
appear in the context of kaon condensation \cite{son,shovkovy},
and gauge symmetry breaking via Bose condensation \cite{sannino}.

\section{one-loop free energy at finite density and temperature}

The bosonic low energy effective Lagrangian for 3D two-color QCD reads
\bea
L&=&\frac{F^2}{2} \tr \nabla_\nu \Sigma(x)\nabla_\nu \Sigma(x)^\dagger
-F^2 M^2 \tr \M \Sigma(x) +L_4.
\label{L}
\eea
where the phenomenological constant $F^2$ has dimensions of mass in 3D, and
$L_4$ is a collection of terms of order $O\( \nabla, M \)^4$.
We take $M\geq0$ without loss of generality.

\subsection{Normal phase }

Within the normal phase (the boundary of which will be determined
in the next section), the Lagrangian reads
\bea
L&=&-4n F^2 M^2
+\frac12 P(x)^\dagger (-\del^2 + M^2) P(x)
+\frac12 Q(x)^\dagger (-\del^2 -4\mu \partial_3 + M^2 -4\mu^2)Q(x)+L_4.
\eea
Thus the one-loop free energy is given by
\bea
\Omega&=&-4n F^2 M^2
+n^2 \tr \log (-\del^2 + M^2)
+n^2 \tr \log (-\del^2 -4\mu \partial_3 + M^2 -4\mu^2)
+L_4[\Sigma=\hat{M}^\dagger]
\nn\\
&=&-4n F^2 M^2
+n^2 \sumint \frac{d^3p}{(2\pi)^3} \log (p^2 + M^2)
+n^2  \sumint \frac{d^3p}{(2\pi)^3}  \log (p^2 +4i \mu p_3 + M^2 -4\mu^2)
+\Omega_4 .
\label{Omeganormal}
\eea
Here $\Omega_4 $ is a constant that is a combination of
the contact term couplings.
The sum-integral in the above is defined as
\be
\sumint d^d p \,f(p_3,{\bf p}) = 
2\pi T \sum_{n=-\infty}^\infty \int d^{d-1} {\bf p}\, f(2n\pi T, {\bf p}).
\ee
The first sum-integral in (\ref{Omeganormal}) is standard
\bea
 -\sumint \frac{d^d p}{(2\pi)^d } \log (p^2 +M^2 )
&=&\frac{1}{(4\pi)^{d/2}}
\int_0^\infty dt \,\frac{\e^{-t M^2}}{t^{d/2+1}}
\sum_{N=-\infty}^\infty \e^{-\frac{n^2}{4tT^2}}\nn\\
&=&\frac{\Gamma(-d/2)}{(4\pi)^{d/2}} M^d
+
4\(\frac{M T}{2\pi}\)^{d/2} \sum_{N=1}^\infty N^{-d/2} K_{d/2}\(\frac{N 
M}{T}\)
\nn\\
& \stackrel{d= 3}{\longrightarrow}&
\frac{M^3}{6\pi} + \frac{T^2}{\pi} \left( M \Li_2(\e^{-\frac{M}{T}}) + T 
\Li_3(\e^{-\frac{M}{T}}) \).
\label{sumintP}
\eea
Note that the zero temperature part is finite in odd dimensional 
spacetime, and the finite temperature correction is expressed in terms 
of the polylogarithm function \cite{abram}
\be
\Li_n(x)=\sum_{k=1}^\infty \frac{x^k}{k^n}.
\ee
The second sum-integral in (\ref{Omeganormal}) is  the
conventional effective action for a bosonic field at finite chemical 
potential,
\bea
&&-\sumint  \frac{d^d p}{(2\pi)^d} \log \((p_3+2i \mu)^2 + {\bf p}^2 + M^2\)
\nn\\
&=&
\frac{1}{(4\pi)^{d/2}} \int_0^\infty
dt \frac{\e^{-t M^2}}{t^{d/2+1}}
\sum_{N=-\infty}^\infty \e^{-\frac{n^2}{4tT^2}}\exp\(\frac{2\mu N}{T} \) 
\nn\\
&=&
\frac{\Gamma(-d/2)}{(4\pi)^{d/2}} M^d +
4\(\frac{M T}{2\pi}\)^{d/2} \sum_{N=1}^\infty N^{-d/2}
K_{d/2}\(\frac{M\, N}{T}\) \cosh\(\frac{2\mu N}{T} \)
\\
& \stackrel{d= 3}{\longrightarrow}&
\frac{M^3}{6\pi} +
\frac{T^2}{2\pi} \left[ M  \( \Li_2(\e^{-\frac{M-2\mu}{T}}) 
+\Li_2(\e^{-\frac{M+2\mu}{T}}) \) +
T  \( \Li_3(\e^{-\frac{M-2\mu}{T}}) +\Li_3(\e^{-\frac{M+2\mu}{T}}) \) 
\right] .
\label{sumintQ}
\eea
Once again, the zero $T$ part is finite due to the odd dimensionality, 
and the finite $T$ corrections are expressed in terms of polylog functions.

Bringing together the contributions (\ref{sumintP}) from the $P$ modes, 
and the contribution (\ref{sumintQ}) from the $Q$ modes, the one-loop 
effective action in the normal phase, at finite $T$ and $\mu$,  is
\bea
\Omega&=&-4n F^2 M^2
-n^2  \left[
\frac{M^3}{6\pi} + \frac{T^2}{\pi} \left( M \Li_2(\e^{-\frac{M}{T}}) + T 
\Li_3(\e^{-\frac{M}{T}}) \)
\right]
\nn\\
&&
-n^2
\left\{
\frac{M^3}{6\pi} +
\frac{T^2}{2\pi} \left[ M  \( \Li_2(\e^{-\frac{M-2\mu}{T}}) 
+\Li_2(\e^{-\frac{M+2\mu}{T}}) \) +
T  \( \Li_3(\e^{-\frac{M-2\mu}{T}}) +\Li_3(\e^{-\frac{M+2\mu}{T}}) \) 
\right]
\right\}
+ \Omega_4 .\label{OmegaNormal}
\eea

\subsection{Diquark condensation phase }

We now compute the free energy,  to 1-loop order,  by expanding about 
the rotated vacuum  \be
\Sigma_\alpha=s_\alpha \M^\dagger s_\alpha^T,\ \
s_\alpha =\exp\(i \frac{\alpha}{2} X_2\)
\label{rotvac}
\ee
which minimizes the tree level effective potential in the diquark 
condensation phase.
We parameterize the fluctuations around $\Sigma_\alpha$ as
\bea
\Sigma(x)&=&s_\alpha S(x) \M^\dagger S(x)^T s_\alpha^T,
\nn\\
S(x)&=&\exp\(\frac{i \Pi(x)}{2F} \),\ \
\Pi(x)=
\frac12
\left[
\ba{cc|cc}
& P &  & Q \\
P^\dagger &  & Q^T &  \\
\hline
& Q^* &  & -P^* \\
Q^\dagger &  & -P^T &
\ea
\right].
\label{SigmaS}
\eea
The quadratic Lagrangian reads
\bea
L&=&-F^2 \mu^2 \tr (\B \Sigma_\alpha\B \Sigma_\alpha^T +\openone_{4n})
-F^2 M^2 \tr \M \Sigma_\alpha
\nn\\
&&+\frac12 P_S(x)^\dagger (-\del^2 + M_1^2 +\frac14 M_3^2) P_S(x)
+\frac12 P_A(x)^\dagger (-\del^2 + M_2^2 +\frac14 M_3^2) P_A(x)
\nn\\
&&
+\frac12 Q(x)^\dagger
\left[
\ba{cc}
-\del^2 + M_1^2 & M_3 \del_3 \\
M_3 \del_3 & -\del^2 + M_2^2
\ea
\right]
Q(x) +L_4.
\eea
Here we have used the notations of \cite{stv1,stv2}, expressing $M_1$, 
$M_2$ and $M_3$ in terms of the overall mass scale $M$, the baryon 
chemical potential $\mu$, and the rotation angle $\alpha$ which 
characterizes the rotated vacuum (\ref{rotvac}) :
\bea
&&M_1^2=M^2 \cos \alpha -4\mu^2 \cos 2\alpha,\nn\\
&&M_2^2=M^2 \cos \alpha -4\mu^2 \cos^2 \alpha,\nn\\
&&M_3^2=16\mu^2 \cos^2 \alpha.
\eea
There are $n(n+1)$  $P_S$ modes, $n(n-1)$ $P_A$ modes, and $(n^2+n^2)$ 
$Q$ modes.
Integrating out these modes, one obtains the one-loop free energy
\bea
\Omega
&=&-F^2 \mu^2 \tr (\B \Sigma_\alpha\B \Sigma_\alpha^T +\openone_{4n})
-F^2 M^2 \tr \M \Sigma_\alpha
\nn\\
&&+\frac12n(n+1) \tr \log (-\del^2 + M_1^2 +\frac14 M_3^2)
+\frac12 n(n-1)\tr \log (-\del^2 + M_2^2 +\frac14 M_3^2)
\nn\\
&&
+\frac12 n^2  \tr \log
\left[
\ba{cc}
-\del^2 + M_1^2 & M_3 \del_3 \\
M_3 \del_3 & -\del^2 + M_2^2
\ea
\right]
+\Omega_4  \nonumber\\
&=&-4n F^2 \(\frac{M_1^2+M^2_2}{2}+ \frac{M_3^2}{4}\)
\nn\\
&&+\frac{n(n+1)}{2}  \sumint \frac{d^3p}{(2\pi)^3} \log (p^2 + M_1^2 
+\frac14 M_3^2)
+\frac{n(n-1)}{2}  \sumint \frac{d^3p}{(2\pi)^3} \log (p^2 + M_2^2 
+\frac14 M_3^2)
\nn\\
&&
+\frac{n^2}{2}  \sumint  \frac{d^3p}{(2\pi)^3} \log
\((p^2 + M_1^2)(p^2 + M_2^2) + p_3^2 M_3^2  \)+\Omega_4 .
\label{dqomega}
\eea
Here $\Omega_4(\a)=L_4[\Sigma_\alpha] $ is given by substituting 
$\Sigma(x)=\Sigma_\alpha$ into
the $O\( \nabla, M \)^4$ terms of the effective Lagrangian.

The $P_S$ and $P_A$ mode contributions in (\ref{dqomega}) are precisely 
as for the $P$ modes in the normal phase in (\ref{sumintP}), with $M$ 
replaced by $\sqrt{M_1^2+M_3^2/4}$ for the $P_S$ modes, and by 
$\sqrt{M_2^2+M_3^2/4}$ for the $P_A$ modes. On the other hand, the final 
term in (\ref{dqomega}) is nontrivial because of the quartic nature of 
the operator. One approach is to factor this quartic as
\bea
(p^2 + M_1^2)(p^2 + M_2^2) + p_3^2 M_3^2  =
(  (p_3+i M y)^2 + {\bf p}^2 + M^2 z^2)( (p_3-i M y)^2 + {\bf p}^2 + M^2 
z^2))-\(\frac{M_1^2-M_2^2}{2}\)^2
\label{qfactor}
\eea
where we have introduced the convenient dimensionless parameters $y$ and 
$z$, as in \cite{stv1,stv2},
\be
y=\frac{M_3}{2M},\ \ \ z=\frac{1}{M}
\sqrt{\frac{M_1^2+M_2^2}{2}+\frac{M_3^2}{4}} .
\ee
One can now expand the  final term in (\ref{dqomega}) in terms of the 
difference
\bea
M_1^2-M_2^2=4\mu^2 \sin^2\alpha .
\label{massdiff}
\eea
Such an expansion proves useful in the Landau-Ginzburg analysis for the 
region where the order parameter $\alpha$ is small, as studied in the 
next Section.

Then the contribution of these $Q$-type modes is
\bea
G_Q&\equiv &
-\sumint  \frac{d^d p}{(2\pi)^d} \log
\((p^2 + M_1^2)(p^2 + M_2^2) + p_3^2 M_3^2  \)
\nn\\
&=&
-\sumint  \frac{d^d p}{(2\pi)^d} \log
\( ( (p_3+i M y)^2 + {\bf p}^2 + M^2 z^2)( (p_3-i M y)^2 + {\bf p}^2 + 
M^2 z^2) \)
\nn\\
&&+\sum_{k=1}^\infty \frac{1}{k} \(\frac{M_1^2-M_2^2}{2}\)^{2k}
\sumint  \frac{d^d p}{(2\pi)^d} \frac{1}{
((p_3+i M y)^2 + {\bf p}^2 + M^2 z^2)^k ( (p_3-i M y)^2 + {\bf p}^2 + 
M^2 z^2)^k }.
\label{kexp}
\eea
The logarithmic term in the second line of (\ref{kexp}) factors in terms of
conventional bosonic effective actions at finite chemical potential :
\bea
G^{(0)}_Q&\equiv &-\sumint  \frac{d^d p}{(2\pi)^d} \log
\left[
( (p_3+i M y)^2 + {\bf p}^2 + M^2 z^2)( (p_3-i M y)^2 + {\bf p}^2 + M^2 
z^2) \right]
\nn\\
&=&
\frac{2}{(4\pi)^{d/2}} \int_0^\infty
dt \frac{\e^{-t (Mz)^2}}{t^{d/2+1}}
\sum_{N=-\infty}^\infty \e^{-\frac{n^2}{4tT^2}}\cosh\(\frac{M y\,N}{T} 
\) \nn\\
&=&
2 \frac{\Gamma(-d/2)}{(4\pi)^{d/2}} (Mz)^d +
8\(\frac{M z T}{2\pi}\)^{d/2} \sum_{N=1}^\infty N^{-d/2}
K_{d/2}\(\frac{M z\, N}{T}\) \cosh\(\frac{M y\,N}{T}\)
\label{k0}\\
& \stackrel{d= 3}{\longrightarrow}&
\frac{(Mz)^3}{3\pi} +
\frac{T^2}{\pi} \left[ M z \( \Li_2(\e^{-M(z-y)/T}) 
+\Li_2(\e^{-M(z+y)/T}) \) +
T  \( \Li_3(\e^{-M(z-y)/T}) +\Li_3(\e^{-M(z+y)/T}) \) \right] .
\nonumber
\eea
The finite-temperature parts of (\ref{k0}) correspond to the corrections 
for masses $M(z\pm y)$.

The $k\geq 1$ terms in the expansion (\ref{kexp}) are more difficult to 
evaluate in a simple closed form. A natural approach is to use Feynman 
parameters:
\bea
&&\sumint  \frac{d^d p}{(2\pi)^d} \frac{1}{
((p_3+i M y)^2 + {\bf p}^2 + M^2 z^2)^k ( (p_3-i M y)^2 + {\bf p}^2 + 
M^2 z^2)^k }
\nn\\
&=&
\frac{\Gamma(2k)}{\Gamma(k)^2}
\sumint  \frac{d^d p}{(2\pi)^d} \int_0^1 ds
\frac{\(s(1-s)\)^{k-1}}{
(p^2 + M^2 (z^2-y^2) +(2s-1)^2iM y p_3 )^{2k} }
\nn\\
&=&
\frac{1}{\Gamma(k)^2(4\pi)^{\frac{d-1}{2}}}
\int_0^1 ds \(s(1-s)\)^{k-1}
\int_0^\infty dt\,t^{2k-1-\frac{d-1}{2}}
\e^{-t M^2( z^2-y^2  )}
T\sum_{n=-\infty}^\infty
\e^{-t [ (2\pi n T)^2 + (2s-1)^2iM y (2\pi n T)]}
\nn\\
&=&
\frac{1}{\Gamma(k)^2(4\pi)^{{d}/{2}}}
\int_0^1 ds \(s(1-s)\)^{k-1}
\int_0^\infty dt\,t^{2k-1-{d}/{2}}
\e^{-t M^2( z^2-y^2  )}
\e^{-t M^2 y^2 (1-2s)^2}
\sum_{N=-\infty}^\infty
\e^{-\frac{n^2}{4tT^2} +\frac{M N}{T}y(1-2s)}
\nn\\
&=&
\frac{2^{3-\frac{d}{2}-4k} \pi^{-\frac{d}{2} }}{\Gamma(k)^2}
\sum_{N=1}^\infty
\(\frac{N}{T M}\)^{2k-\frac{d}{2}}
\int_0^1 \frac{du\,u^{k-1}}{\sqrt{1-u}}
\frac{K_{\frac{d}{2}-2k} \( \frac{M N}{T}\sqrt{z^2-y^2 u}\) }{(z^2-y^2 
u)^{k-\frac{d}{4}}}
\cosh\( \frac{M N}{T}y\sqrt{1-u} \)+ \ \mbox{($T=0$ part)}
\nn\\
&\stackrel{d=3}{\longrightarrow}&
\frac{2^{\frac{3}{2}-4k} \pi^{-\frac{3}{2} }}{\Gamma(k)^2}
\sum_{N=1}^\infty
\(\frac{N}{T M}\)^{2k-\frac{3}{2}}
\int_0^1 \frac{du\,u^{k-1}}{\sqrt{1-u}}
\frac{K_{\frac{3}{2}-2k} \( \frac{M N}{T}\sqrt{z^2-y^2 u}\) }{(z^2-y^2 
u)^{k-\frac{3}{4}}}
\cosh\( \frac{M N}{T}y\sqrt{1-u} \)+ \ \mbox{($T=0$ part)} .
\nn\\
\label{kterms}
\eea

Motivated by the Landau-Ginzburg analysis of the next Section, we 
consider just  the $k=1$ term, since higher terms are at least of order 
$\alpha^8$. In $d=3$ the $k=1$ term in (\ref{kterms}) reduces to a simple 
integral :
\bea
G^{(1)}_Q&\stackrel{d= 3}{\longrightarrow}&
\frac{1}{8\pi M y}\left\{ \frac12 \log \frac{z+y}{{z-y}} +
\int^{(z+y)M/T}_{(z-y)M/T}  \frac{dx}{x(\e^{x}-1)}\right\} .
\label{22}
\eea
Here the first term corresponds to the $T=0$ contribution, while the 
integral represents the finite $T$ correction.

Collecting together the contributions of the $P_S$, $P_A$ and $Q$ modes, 
we obtain the following expression for the one-loop effective action in 
the diquark phase, at finite temperature and chemical potential
\bea
&&\Omega=-4nF^2 M^2 z^2
\nn\\
&&-\frac{n(n+1)}{2}
\left[\frac{(M_1^2+M_3^2/4)^{3/2}}{6\pi} + \frac{T^2}{\pi}
\left( \sqrt{M_1^2+{M_3^2\over4}}
\Li_2(\e^{-\sqrt{M_1^2+M_3^2/4}/T}) + T 
\Li_3(\e^{-\sqrt{M_1^2+M_3^2/4}/T}) \)
\right]
\nn\\
&&-\frac{n(n-1)}{2}  \left[\frac{(M_2^2+M_3^2/4)^{3/2}}{6\pi} + 
\frac{T^2}{\pi}
\left( \sqrt{M_2^2+{M_3^2\over4}}
\Li_2(\e^{-\sqrt{M_2^2+M_3^2/4}/T}) + T 
\Li_3(\e^{-\sqrt{M_2^2+M_3^2/4}/T}) \)
\right]
\nn\\
&&-\frac{n^2}{2}
\left\{
\frac{M^3z^3}{3\pi} +
\frac{T^2}{\pi} \left[ M z \( \Li_2(\e^{-(z-y)M/T}) 
+\Li_2(\e^{-(z+y)M/T}) \) +
T  \( \Li_3(\e^{-(z-y)M/T}) +\Li_3(\e^{-(z+y)M/T}) \) \right]
\right.
\nn\\
&&
\left.+
\frac{(M_1^2-M_2^2)^2}{32\pi y M}
\left[\frac12  \log \frac{z+y}{z-y} +\int^{(z+y)M/T}_{(z-y)M/T}  
\frac{dx}{x(\e^{x}-1)}
\right]
\right\}
+O(M_1^2-M_2^2)^4 +\Omega_4.
\label{Omega}
\eea
Note that the higher order corrections are at least of order 
$(M_1^2-M_2^2)^4\sim \mu^8 \, \sin^8\alpha$.

\section{Landau-Ginzburg model}

In the Landau-Ginzburg approach \cite{landau}, we regard the rotation 
angle $\alpha$ as an order parameter. For small $\alpha$, the shape of 
the effective potential can be used to deduce information about the 
phase structure of the system \cite{huang}. The vanishing of the 
coefficient of the term quadratic in $\alpha$ determines a critical line 
in the $T$ and $\mu$ plane.

\subsection{Landau-Ginzburg model at zero temperature}

It is important first to analyze the Landau-Ginzburg model at $T=0$ in 
order to understand the effect of one-loop renormalization. Since we 
expect the critical behavior to be near the tree-level $T=0$ critical 
point where $\mu=M/2$, we expand $\Omega$ in terms of the rotation angle 
$\alpha$ and also the deviation of chemical potential from $M/2$:
\be
\bm\equiv\frac{\mu}{M}-\frac12 .
\ee
Substituting the definitions of $M_1$, $M_2$, $M_3$, $z$ and $y$ into 
the $T=0$ parts of the one-loop effective potential in (\ref{Omega}) and 
expanding, one finds
\be
\frac{\Omega_{T=0}}{-4nF^2 M^2}=
{\rm cst.}+\frac{1}{32\pi} \frac{M}{F^2} \a^2
+ \( 2 + \frac{2n+1}{8\pi } \frac{M}{F^2} \) \bm \a^2+
O\(\a^4 \log(\a^2-4\bar{\mu})\).
\label{lg0}
\ee
Note that the higher order terms in (\ref{Omega}) are not relevant as 
they contribute terms at least of order $\alpha^8$.
Also, we neglect the quartic contact terms $\Omega_4$.
This is motivated by the 4D case \cite{stv2}
where these terms are negligibly small.
The dimensionless combination $M/F^2$ in (\ref{lg0}) is
the loop expansion parameter.
The vanishing of the coefficient of $\a^2$ leads to the
critical point (at one-loop)
\bea
\bm&=&-\frac{1}{64\pi} \frac{M}{F^2},
\label{mubar}
\eea
This can be rewritten as
\bea
\mu &=&\frac{1}{2}M\(1- \frac{1}{32\pi}\frac{M}{F^2} \)
\label{shift}
\eea
which shows that the critical point is shifted by a finite one-loop 
correction from the tree-level $T=0$ critical point $\mu=M/2$. Actually, 
this shift is just a  mass renormalization since the one-loop 
renormalized pion mass $m_{\pi}$ is
\be
m_{\pi}=
M \(1+\frac{\Gamma(1-d/2)}{8(4\pi)^{d/2}}\frac{M^{d-2}}{F^2} \)
\stackrel{d= 3}{\longrightarrow}
M\(1- \frac{1}{32\pi}\frac{M}{F^2} \)
\label{mpi}
\ee
as computed in the Appendix. Thus, this result means that  one-loop 
renormalization has the effect of shifting the $T=0$ critical point from 
$\mu=M/2$ to $\mu=m_\pi/2$, where $m_\pi$ is the one-loop renormalized 
value of $M$.

\subsection{Landau-Ginzburg model at finite temperature}

At finite temperature, the shape of the free energy as a function of the 
order parameter $\alpha$ changes as we move through the $T$ and $\mu$ 
plane. Given the explicit expression (\ref{Omega})
it is straightforward to plot (using, {\it e.g.}, Mathematica) the free 
energy for chosen values of $T$ and $\mu$. This gives some guidance as 
to the parameter region in which to search for a critical line. We find 
that the appropriate critical region of parameters is
\be
0<\mu -\frac{m_\pi}{2} \simeq   m_\pi  g\ll {T} \ll m_\pi  ,
\label{critregion}
\ee
where we have defined the natural dimensionless
self coupling constant of pions by
\be
g=\frac{M}{32\pi F^2} .
\ee
In this region
the temperature is far smaller than the mass of
massive modes (the $P$'s and half of the $Q$'s), but
is far larger than the mass of the almost Goldstone modes
(the other half of the $Q$'s).
We also have the intrinsic applicability limit of the low energy 
effective Lagrangian that
other hadronic modes are not excited:
i.e., $m_\pi \ll \Lambda_{\rm QCD_3}$.

We expand the Laudau-Ginzburg free energy
(normalized by that at $\mu=T=0$)
up to the quartic order in $\alpha$,
\be
\frac{\Omega}{-4n F^2 M^2}=c_0+c_2 \alpha^2 +c_4 \alpha^4+O(\alpha^6).
\label{lgexp}
\ee
Using the expression (\ref{Omega}), the coefficient of the quadratic term is given explicitly by
\bea
&&c_2=
-\frac12+\frac{2\mu^2}{M^2}+
\frac{1}{32\pi F^2 M^2}
\left[
-2n M^3 + 4( 1 + 2n)  \mu^2 M
+ 2( -n M^2  + 4 ( 1 + n)\mu^2 ) T\log (1 - \e^{-\frac{M}{T}})
\right.
\nn\\
&&
\left.
+ ( -M^2 + 4\mu M+ 4\mu^2 ) n T \log (1 - \e^{-\frac{M - 2\mu}{T}})
+ ( -M^2 - 4\mu M+ 4\mu^2 ) n T \log (1 - \e^{-\frac{M + 2\mu}{T}})
\right.
\nn\\
&&
\left.
-  4n \mu   T^2 \left(
\Li_2(\e^{-\frac{M - 2\mu}{T}}) - \Li_2(\e^{-\frac{M + 2\mu}{T}})
\right)
\right] .
\label{c2exact}
\eea
Similarly, from (\ref{Omega}) the coefficient of the quartic term is given explicitly by
\bea
&&c_4= \frac{1}{24} -\frac{\mu^2}{3M^2}
+ \frac{1}{768\pi F^2 M^2}
\left[
10n M^3 - 56(1+2n)M\mu^2 + 48(2+3n) \frac{\mu^4}{M}
\right.\nn\\ &&\left.
+ \frac{3n(M^2 - 4M\mu  - 4\mu^2)^2}{(\e^{\frac{M- 2\mu}{T}}-1 ) M}
+ \frac{3n(M^2 + 4M\mu  - 4\mu^2)^2}{(\e^{\frac{M + 2\mu}{T}}-1) M}
+  \frac{6\left( n M^4 - 8(1+ n)M^2\mu^2 + 32(1+n) \mu^4 \right)}{( 
\e^{\frac{M}{T}}-1 ) M}
\right.\nn\\ &&\left.
+  2n ( M^2 - 4M\mu - 40\mu^2) T\log (1 - \e^{-\frac{M - 2\mu}{T}})
+  2n (M^2 + 4M\mu  - 40\mu^2) T\log (1 - \e^{-\frac{M + 2\mu }{T}})
\right.\label{c4exact}\\ &&\left.
+  4\left( n M^2 - 16(1+n)\mu^2 \right) T\log (1 - \e^{-\frac{M}{T} })
+  8n  \mu T^2 \left( \Li_2(\e^{-\frac{M - 2\mu}{T}}) - 
\Li_2(\e^{-\frac{M + 2\mu}{T}}) \right)
+ 12n \mu^3\int_{\frac{M - 2\mu}{T}}^{\frac{M + 
2\mu}{T}}\frac{dx}{x}\coth \frac{x}{2}
\right] .
\nn
\eea
The separation between the tree level part and the one-loop part
is apparent in these expressions for $c_2$ and $c_4$.
We stress that no approximation has been used so far, except that 
(\ref{lgexp}) is an expansion for small values of the order parameter 
$\alpha$.

In the parameter region (\ref{critregion}) we may obtain more compact approximate
expressions for $c_2$ and $c_4$ by the use of
the expansion formula of polylogarithms with positive integer indices,
\be
\Li_n(\e^{-\epsilon})=
\frac{\psi(n)+\gamma-\log \epsilon}{(n-1)!} (-\epsilon)^{n-1}
+\sum_{k\geq 0 \atop k\neq n-1} \frac{\zeta(n-k) }{k!}(-\epsilon)^k .
\ee
and the formula
\be
\int_\epsilon^{\Lambda} \frac{dx}{x} \coth\frac{x}{2} =
\frac{2}{\epsilon}+ \log \Lambda 
+O(\epsilon^0)+O\(\frac{\e^{-\Lambda}}{\Lambda}\).
\ee
valid for $0<\epsilon \ll 1 \ll \Lambda$.
We define the dimensionless deviation parameter
\be
\delta=1-\frac{2\mu}{M}\equiv -2\bar{\mu}
\ee
which is of the same infinitesimal order as the coupling $g$.
Then we find the following simple approximate formulas for $c_2$ and $c_4$:
\bea
c_2&\simeq&
-\delta+g\(1-2n\frac{T}{M} \log\frac{T}{M\, \delta } \) ,\\ \nn\\
c_4&\simeq& -\frac{1}{24}  +g  \frac{5n}{24} \frac{T}{M \,\delta} .
\label{c2c4approx}
\eea
In the Landau-Ginzburg approach, the critical line is characterized by 
the condition $c_2=0$, and the tricritical point is characterized by the 
conditions $c_2=c_4=0$ \cite{huang}.
Expressing dimensionful quantities
in the unit of the renormalized pion mass $m_\pi$,
rather than the bare pion mass $M$,
the  critical conditions given by the approximated coefficients
(\ref{c2c4approx}) read
\bea
\mbox{critical line:}&&
\mu-\frac{m_\pi}{2}=
n g T \log \frac{T}{ m_\pi -2\mu +g m_\pi} ,
\label{critline}\\ \nn\\
\mbox{tricritical point:}&&
\mu_{\rm tri}=
\frac{m_\pi}{2}\(
1+\frac{g}{1-\frac{5}{2\log(5ng)}} \),\ \
T_{\rm tri}=
\frac{m_\pi}{n\(5-2\log(5ng)\)} .
\label{tricritpoint}
\eea
We have confirmed numerically that these expressions approximate very 
well the critical line and tricritical point derived numerically from 
the vanishing of the full expressions for $c_2$ and $c_4$ in 
(\ref{c2exact},\ref{c4exact}), for the parameter region of concern.
Accordingly our approximation is indeed self-consistent
in that the critical region is within the region (\ref{critregion}).

\begin{figure}
\includegraphics{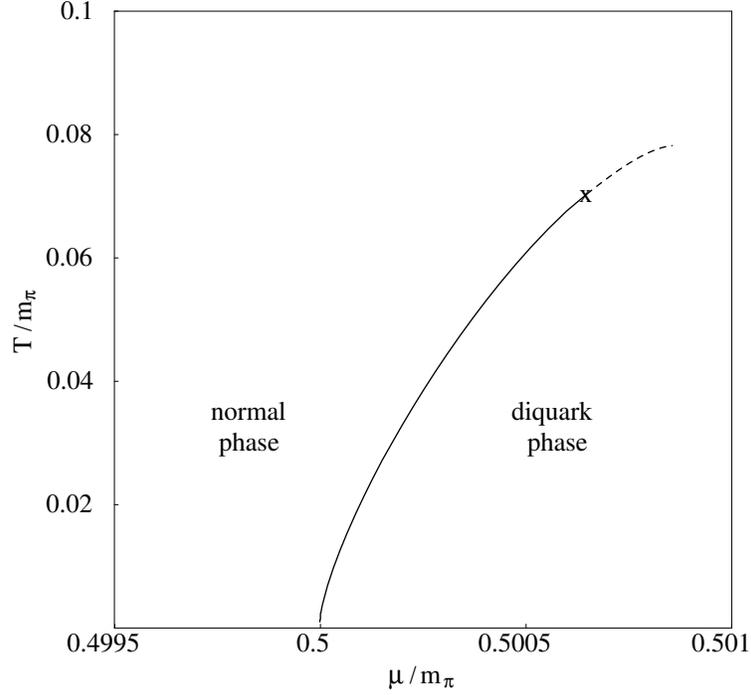}
\caption{Plot of the phase diagram for two flavors ($n=1$)
and coupling $g=0.002$
which corresponds to $M/F^2=0.2$. 
The solid line is the critical line
in (\ref{critline}), and the x marks the tricritical point (\ref{tricritpoint}).
The dotted line is an approximation based on expanding the effective
potential to sixth order in $\alpha$ and demanding it to have the form 
$\Omega(\alpha)=c_6 \alpha^2 (\alpha^2-{\rm const.})^2+{\rm const.}$
Note that the scale has been magnified to show
just the vicinity of the critical line.}
\end{figure}

\subsection{The critical line and a weakly interacting Bose gas}

The critical line (\ref{critline}) has an interpretation in terms of a 
weakly interacting Bose gas, as in the $3+1$ dimensional case studied in 
\cite{stv1,stv2}, where the diquark phase was argued to be a Bose 
condensate. The fact that the same structure appears in our $2+1$ 
dimensional model is quite interesting, as Bose condensation is rather 
different in two spatial dimensions. So we show here explicitly how the 
usual Bose gas results are consistent with the Landau-Ginzburg effective 
Lagrangian approach of this paper.

Recall that for a free relativistic Bose gas in $d$ spatial dimensions, 
the charge density is given by \cite{haber}
\begin{eqnarray}
\rho=\frac{\Gamma\left(\frac{d+1}{2}\right)}{\pi^{(d+1)/2}}\, T^d\, 
\left[ g_d\left(\frac{M}{T},\frac{\mu}{M}\right)- 
g_d\left(\frac{M}{T},-\frac{\mu}{M}\right)\right] .
\label{rhod}
\end{eqnarray}
The first term corresponds to the particles and the second to the 
antiparticles. For $d=2$ \cite{haber},
\begin{eqnarray}
g_2\left(\frac{M}{T},\frac{\mu}{M}\right)=
\Li_2\left(\e^{(\mu-M)/T}\right)-\frac{M}{T}\,\log\left(1- 
\e^{(\mu-M)/T}\right) .
\label{g2}
\end{eqnarray}

In the $2+1$ dimensional two-color QCD system discussed in this paper, 
the $Q$ excitations have baryon charge $2$, so we should replace the 
chemical potential in (\ref{rhod},\ref{g2}) by $\mu\to 2\mu_B$, and the 
charge density in (\ref{rhod}) by $\rho\to n_B/2$.  At low temperature 
the lightest modes, the $Q$ modes, dominate. There are
$n^2$ such modes, where we recall that the number of flavors is $N_f=2n$.
Furthermore, at low temperature only the particles contribute and so we 
expect
\begin{eqnarray}
n_B=2 n^2 \frac{T^2}{2\pi}\, g_2\left(\frac{M}{T},\frac{2\mu_B}{M}\right) .
\label{nbose}
\end{eqnarray}
In the Landau-Ginzburg approach discussed in the previous section, the 
critical line separating the normal and diquark phases is determined by 
the vanishing of the coefficient $c_2$ of the term quadratic in the 
order parameter $\alpha$ in the expansion of the free energy 
(\ref{lgexp}). At one-loop, the vanishing of $c_2$ in (\ref{c2exact}) 
near the critical point $\mu_B \approx m_\pi/2$  can be written as [here 
we use slightly less restrictive approximations than were used in 
obtaining (\ref{critline})]
\begin{eqnarray}
\mu_B-\frac{m_\pi}{2}&\approx&  \frac{n T^2}{32 \pi F^2} 
\left[\Li_2\left(\e^{(2\mu_B-M)/T}\right)-\frac{M}{T}\,\log\left(1- 
\e^{(2\mu_B-M)/T}\right)\right]\nonumber\\
&=& \frac{n T^2}{32\pi F^2}\, g_2\left(\frac{M}{T},\frac{2\mu_B}{M}\right) .
\label{lg}
\end{eqnarray}
Combining these two results (\ref{nbose}) and (\ref{lg}) we see that
\begin{eqnarray}
n_B=16 N_f F^2 \left(\mu_B-\frac{m_\pi}{2}\right).
\label{nmu}
\end{eqnarray}
In the effective Lagrangian approach the renormalized vacuum energy is 
\footnote{Note that in (\ref{energy}) we differ from \cite{stv1,stv2}  
by a factor of 2 in the interaction term. This is because only half of 
the $Q$ modes  contribute to the low $T$ Bose gas, and so only they 
should be selected out of the general $\Pi$ field in this limit.}
\begin{eqnarray}
{\cal E}_{\rm vac}=n_B\, \frac{m_\pi}{2}+\frac{n_B^2}{32 F^2 N_f}+\cdots .
\label{energy}
\end{eqnarray}
where the coefficients are fixed by the symmetries.
Since the baryon number chemical potential is
\begin{eqnarray}
\mu_B=\frac{\partial {\cal E}_{\rm vac}}{\partial n_B} 
\end{eqnarray}
we see that the relation (\ref{nmu}) follows directly. This demonstrates 
the consistency of the Landau-Ginzburg effective Lagrangian approach 
with the standard Bose gas  results, and shows that the critical line 
separating the normal and diquark phases describes a free Bose gas in 
two spatial dimensions.

\section{Conclusion and Discussion}
In this paper we have used the low energy effective field theory method 
at finite temperature and finite baryon density to investigate the phase 
structure of three dimensional parity invariant
SU(2) QCD with fundamental quarks. We computed the one-loop effective 
potential at both $T=0$ and finite $T$, at nonvanishing baryon number 
chemical potential. At $T=0$, the tree-level critical point $\mu=M/2$ 
between the normal and diquark phase receives a finite shift at 
one-loop, but this shift corresponds exactly to the mass 
renormalization, so that at one-loop the $T=0$ critical point is 
$\mu=m_\pi/2$, where $m_\pi$ is the one-loop renormalized pion mass.
At nonzero temperature we found a simple expression (\ref{Omega}) for 
the effective potential, suitable for a Landau-Ginzburg analysis in 
terms of the order parameter $\alpha$ which describes the rotation of 
the tree-level vacuum away from the normal phase vacuum alignment. The 
subsequent Landau-Ginzburg analysis of this free energy revealed a 
critical line in the $(\mu, T)$ plane, given approximately by 
(\ref{critline}). This line
separates the normal and diquark phases, and ends in a tricritical 
point whose location is given approximately by (\ref{tricritpoint}). 
This general  structure, with a critical line ending in a tricritical 
point, is qualitatively similar to that found in the 4D case 
\cite{stv2}, except that the $T\, \log T$ temperature dependence for the 
critical line in (\ref{critline}) becomes a power law $\sim T^{3/2}$ in 
the 4D case.

However, the details of the 3D phase diagram are much less clear
than in the 4D case, due to the presence of infrared divergences, which 
become more pronounced at higher loops, where they will appear as powers.
In a restricted region of the relevant coupling and parameters (such as 
$g$, $T/M$ and $\bar{\mu}$), the one-loop effective potential does 
describe the correct physics.
The one-loop calculation indicates
that some transition happens in the vicinity of the critical line 
(\ref{critline}),
and the one-loop Landau-Ginzburg analysis suggests that this is a line 
of second-order phase transitions, as is the case in 4D \cite{stv2}. 
However, in 3D there are infrared divergences as $\bar{\mu}$ tends to 
zero ($\mu\to M/2$), which become more significant at higher loops. This 
suggests that a nonperturbative analysis may be necessary to identify 
the true nature of the phase transition. There is another reason to 
think that in the 3D case
the critical line may not be one of second-order phase transitions:
in two spatial dimensions, only truly massless modes Bose condense -- 
massive modes, no matter how light, do not Bose condense \cite{haber}.
If the $\tilde{Q}$ modes are truly massless even at nonzero temperature,
we would expect
to be able to dimensionally reduce the system
 to two dimensions, in which case the Mermin-Wagner-Coleman theorem 
tells us that a continuous symmetry cannot be broken
(in the absence of long-range interactions). What then is the fate of 
the diquark phase at nonzero $T$? Perhaps nonperturbative effects generate
an exponentially small, $\sim \e^{-F^2/T}$ (rather than actually zero), 
mass for the modes (which at tree level are exactly massless), and the 
critical line is actually a cross-over between a normal and diquark 
phase, until the tricritical point is reached, after which it presumably 
becomes a line of first-order transitions.
 We mention that such a nonperturbative exponentially small mass has 
been observed in the thermodynamics of the
 large $N$ $O(N)$ sigma model in 3D \cite{baruch}.
This would mean that the standard perturbative effective potential approach
to two-color QCD in 3D is good for describing the physics on some 
portion of the critical line where it is a noninteracting Bose gas, but 
away from this region  the perturbative analysis is insufficient.
It would be very interesting to investigate the relevant nonperturbative effects, 
perhaps from a large $N_F$ analysis, or from a lattice study of 3D 
two-color QCD.

\begin{acknowledgments}
We thank A. Kovner,  V. Miransky, I. Shovkovy, K. Splittorff  
and J. Verbaarschot for helpful discussions. We also thank the U.S. DOE 
for support through the grant DE-FG02-92ER40716.
\end{acknowledgments}

\appendix
\section{One-Loop Renormalization}
In this Appendix  we perform the one-loop renormalization of the
pion mass and wave function. As the theory is renormalizable at each 
order of the chiral perturbation, renormalization constants are common 
for both
normal and diquark phase. Thus we shall concentrate on the simpler
case of the normal phase, and set $\Omega_4=0$ as mentioned in section IV.A.
We substitute (\ref{SigmaS}) at $\alpha=0$ into the tree-level term
of (\ref{L}) and expand
in $P$ and $Q$ fields up to the quartic order.
We write
\be
P_{ab}=\phi^{(1)}_{ab} + i \phi^{(2)}_{ab} ,\ \
Q_{ab}=\phi^{(3)}_{ab} + i \phi^{(4)}_{ab} ,\ \
a,b=1,\ldots,n .
\ee
Without loss of generality we choose to compute
$\< \phi_{11}^{(1)}(p) \phi_{11}^{(1)}(-p)\>$.
The terms that contribute to this Green's function are:
\bea
L_{\rm tree}&=&\sum_{\alpha=1}^4 \sum^n_{a,b=1}
\frac12 \phi_{ab}^{(\alpha)}\( -\partial^2+M^2\) \phi_{ab}^{(\alpha)}
\nn\\
&-&\frac{M^2}{96F^2}
\left\{ (\phi_{11}^{(1)})^4+
2(\phi_{11}^{(1)})^2 \left[
\sum_{\alpha=2}^4  (\phi_{11}^{(\alpha)})^2  +
\sum_{\alpha=1}^4 \sum_{a=1}^n
\(  (\phi_{a1}^{(\alpha)})^2  +  (\phi_{1a}^{(\alpha)})^2 \)
\right]\right\}
\nn\\
&-&\frac{1}{96F^2}
(\phi_{11}^{(1)})^2 \left[
4\sum_{\alpha=2}^4 \  (\partial \phi_{11}^{(\alpha)})^2  +
\sum_{\alpha=1}^4 \sum_{a=1}^n
\(  (\partial\phi_{a1}^{(\alpha)})^2  +  (\partial\phi_{1a}^{(\alpha)})^2 \)
\right]
\nn\\
&-&\frac{1}{96F^2}
(\partial\phi_{11}^{(1)})^2 \left[
4\sum_{\alpha=2}^4  ( \phi_{11}^{(\alpha)})^2  +
\sum_{\alpha=1}^4 \sum_{a=1}^n
\(  (\phi_{a1}^{(\alpha)})^2  +  (\phi_{1a}^{(\alpha)})^2 \)
\right] . 
\eea
There exist vertices of the form
$\phi_{11}^{(1)} \partial_\nu \phi_{11}^{(1)}
\phi_{1a}^{(\alpha)} \partial_\nu \phi_{1a}^{(\alpha)}, a\neq 1$,
but they do not contribute
to $\< \phi_{11}^{(1)}\phi_{11}^{(1)}\>$.

Using these vertices, we have
\bea
&&\< \phi_{11}^{(1)}(p) \phi_{11}^{(1)}(-p)\> \nn\\
&=&
\frac{1}{(p^2+M^2)^2}
\left[p^2+M^2
+\frac{M^2}{96F^2} 4\cdot 3
\int\frac{d^d q}{(2\pi)^d} \frac{1}{q^2+M^2}
+\frac{M^2}{48F^2}   2\cdot 1(3+8(n-1))
\int\frac{d^d q}{(2\pi)^d} \frac{1}{q^2+M^2}
\right. \nn\\
&&
\left.
-\frac{1}{96F^2}   2\cdot 1(12+8(n-1))
\int\frac{d^d q}{(2\pi)^d} \frac{-q^2}{q^2+M^2}
-\frac{1}{96F^2}2\cdot 1 (12+8(n-1)) (-p^2)
\int\frac{d^d q}{(2\pi)^d} \frac{1}{q^2+M^2} \right]
\nn\\
&=&
\frac{1}{(p^2+M^2)^2}
\left[
p^2\(1+ \frac{2n+1}{12F^2} \int\frac{d^d q}{(2\pi)^d} \frac{1}{q^2+M^2} \)
+
M^2 \(1+\frac{n-1}{6F^2} \int\frac{d^d q}{(2\pi)^d} \frac{1}{q^2+M^2}  \)
\right]
\nn\\
&=&
\frac{1}{(p^2+M^2)^2}
\left[
p^2\(1+ \frac{2n+1}{12F^2} \frac{\Gamma(1-d/2)}{(4\pi)^{d/2}} M^{d-2} \)
+
M^2 \(1+\frac{n-1}{6F^2}\frac{\Gamma(1-d/2)}{(4\pi)^{d/2}} M^{d-2} \)
\right].
\label{A3}
\eea
We have discarded a term proportional to the divergent integral
$\int\frac{d^d q}{(2\pi)^d} $
in the dimensional regularization.

The wave function renormalization should absorb the factor
multiplying $p^2$,
\be
\phi(p)= \phi(p)_{\rm ren}
\(1+ \frac{2n+1}{24F^2} \frac{\Gamma(1-d/2)}{(4\pi)^{d/2}} M^{d-2} \).
\ee
Accordingly,
\be
\< \phi_{11}^{(1)}(p){}_{\rm ren} \phi_{11}^{(1)}(-p){}_{\rm ren}\>=
\frac{1}{(p^2+M^2)^2}
\left[
p^2 +
M^2 \(1-\frac{1}{4F^2}\frac{\Gamma(1-d/2)}{(4\pi)^{d/2}} M^{d-2} \)
\right].
\ee
At $p=0$,
\be
\< \phi_{11}^{(1)}(0){}_{\rm ren} \phi_{11}^{(1)}(0){}_{\rm ren}\>=
\frac{1}{M^2}
\(1-\frac{\Gamma(1-d/2)}{4(4\pi)^{d/2}} \frac{M^{d-2}}{F^2} \)
\equiv\frac{1}{m_{\pi}^2} .
\ee
That is,
\be
m_{\pi}= M\(1+\frac{\Gamma(1-d/2)}{8(4\pi)^{d/2}}\frac{M^{d-2} }{F^2} \) .
\ee
Note that the $n$-dependent part of the
coefficients of $p^2$ and $M^2$ in (\ref{A3}) are identical, leading to
the $n$-independent mass renormalization.

\end{document}